\documentclass[preprint,amsmath,amssymb,aps,longbibliography]{revtex4-1}
\usepackage{graphicx}
\usepackage{dcolumn}
\usepackage[normalem]{ulem}
\usepackage{bm}
\usepackage[utf8]{inputenc}
\usepackage[T1]{fontenc}
\usepackage{mathptmx}
\usepackage{nth}
\usepackage{cases}
\usepackage{xcolor}

\newcommand\bracket[1]{\mathbin{\left\langle #1 \right\rangle}}
\newcommand\brackett[1]{\mathbin{\langle #1 \rangle}}
\DeclareRobustCommand{\bchi}{{\mathpalette\irchi\relax}}
\newcommand{\irchi}[2]{\raisebox{\depth}{$#1\boldsymbol{\chi}$}}

\newcommand{\ee}{\mathbin{\mathrm{e}}}
\begin{document}

\preprint{APS/123-QED}

\title[Dispersion and reaction in random flows]{Dispersion and reaction in random flows:\\single realisation vs ensemble average}

\author{Antoine Renaud}
 \email{antoine.renaud@ed.ac.uk}
\author{Jacques Vanneste}
\affiliation{School  of  Mathematics  and  Maxwell  Institute  for  Mathematical  Sciences,  University  of  Edinburgh, King’s Buildings, Edinburgh EH9 3FD, United Kingdom }
\date{\today}

\begin{abstract}
We examine the dispersion of a passive scalar released in an incompressible fluid flow in an unbounded domain. 
The flow is assumed to be spatially periodic, with zero spatial average, and random in time, in the manner of the random-phase alternating sine flow which we use as an exemplar. In the long-time limit, the scalar concentration takes the same, predictable form for almost all realisations of the flow, with a Gaussian core characterised by an effective diffusivity, and large-deviation tails characterised by a rate function (which can be evaluated by computing the largest Lyapunov exponent of a family of random-in-time partial differential equations). We contrast this single-realisation description with that which applies to the average of the concentration over an ensemble of flow realisations. We show that the single-realisation and ensemble-average effective diffusivities are identical but that the corresponding rate functions are not, and that the ensemble-averaged description overestimates the concentration in the tails compared with that obtained for single-flow realisations. This difference has a marked impact for scalars reacting according to the Fisher--Kolmogorov--Petrovskii--Piskunov (FKPP) model. Such scalars form an expanding front whose shape is approximately independent of the flow realisation and can be deduced from the single-realisation large-deviation rate function. We test our predictions against numerical simulations of the alternating sine flow.

\end{abstract}

\maketitle

\section{\label{sec:Intro}Introduction}

We are interested in the dispersion of a passive scalar field under the combined action of advection by a prescribed flow and molecular diffusion. The scalar concentration $C\left(\bm{x},t\right)$ is governed by the advection--diffusion equation 
\begin{equation}\label{eq:C_advdiff}
    \partial_{t}C+\bm{u}\cdot\nabla C=\kappa\nabla^2 C,
\end{equation}
where  $\kappa$ is the molecular diffusivity and $\bm{u}\left(\bm{x},t\right)$ a velocity field satisfying $\nabla\cdot\bm{u}=0$ as appropriate for incompressible flows. In the Lagrangian interpretation of \eqref{eq:C_advdiff}, the positions ${\bm X}(t)$ of scalar particles obey the stochastic differential equation 
\begin{equation}\label{eq:SDE}
    \mathrm{d}\bm{X}=\bm{u}\left(\bm{X}(t),t\right)\mathrm{d} t+\sqrt{2\kappa} \, \mathrm{d}\bm{W},
\end{equation}
where ${\bm W}$ is a multidimensional Brownian motion. When suitably normalised, the concentration in \eqref{eq:C_advdiff} is identified with the probability density function
\begin{equation}\label{eq:WtoC}
    C\left(\bm{x},t\right)=\mathbb{E}\left[\delta(\bm{X}\left(t\right)-\bm{x}\right)],
\end{equation}
of the particle positions, where $\mathbb{E}$ denotes expectation over realisations of the Brownian motion conditioned on ${\bm{X}(0)}={\bm 0}$ so that $C({\bm x},0)=\delta({\bm x})$. 

We focus on random flows, taking ${\bm u}({\bm x},t)$ to be a random field with given stationary statistics, as is standard when modelling turbulent dispersion
\cite{taylor1922,kraichnan1966,majda1999,falkovich2001}. This introduces an ensemble of flow realisations, additional to the ensemble of Brownian motion, with an associated average which we denote by $\bracket{\cdot}$. The averaged scalar concentration, defined as
\begin{equation} \label{eq:WtoCav}
\brackett{C \left(\bm{x},t\right)}=\brackett{\mathbb{E}\left[\delta(\bm{X}\left(t\right)-\bm{x}\right]},
\end{equation}
is often the main object of investigation \cite[e.g.][]{majda1999}. However, for practical purposes, when a single flow realisation is experienced, it is the single-realisation dynamics of $C$ that matters. For large times and/or on large spatial scales, ergodicity arguments can be invoked to identify $C$ with $\brackett{C}$. The relationship between the single-realisation $C$ and the ensemble-averaged $\brackett{C}$ is in fact rather subtle; it is the topic of this paper. In statistical-physics parlance, the fixed flow realisation associated with $C$ corresponds to a  quenched disorder, while the sampling over realisations associated with $\brackett{C}$ corresponds to an annealed disorder.

We investigate the relationship between $C$ and $\brackett{C}$ in the context of the initial-value problem for \eqref{eq:C_advdiff} in an unbounded domain, corresponding to the instantaneous release of a localised patch of scalar. In the absence of advection (${\bm u}=0$), the concentration  behaves asymptotically as the Gaussian $(4\pi\kappa t)^{-d/2}\ee^{-|\bm{x}|^2/\left(4\kappa t\right)}$, with $d$ the number of spatial dimensions. For a spatially periodic, time-independent ${\bm u} \not= 0$, at a coarse-grained level, the core of the concentration is also Gaussian: the dynamics of $C$ is approximately diffusive, with a (tensorial) effective diffusivity which is computed through the solution of the cell problem of homogenisation theory (see, e.g., \cite{majda1999,PavliotisStuart2007} and references therein). The tails, however, are non-Gaussian: they take a large-deviation form, characterised by a rate function which is deduced from the solution of an eigenvalue problem (generalising the cell problem of homogenisation) and encodes all the cumulants of the concentration \cite{HaynesVanneste2014}. These conclusions extend naturally to flows that are also periodic in time. 

In this paper, we examine the case of spatially-periodic, random-in-time flows. This class includes the widely-studied random alternating sine flow introduced by \citet{Perrehumbert1994} which we use as our testbed. For these flows, the coarse-grained picture of a Gaussian core and large-deviation tails continues to hold, with a single effective diffusivity and a single rate function describing the concentration $C$ in almost all realisations of the flow. On the other hand, the ensemble-averaged concentration $\brackett{C}$ also has a Gaussian core and large deviations tails. We show that, under the assumption of a vanishing spatial average of ${\bm u}$, the Gaussian cores of $C$ and $\brackett{C}$ coincide; in other words, the single-realisation and ensemble-average effective diffusivities are identical. The large-deviation tails of $C$ and $\brackett{C}$ are generally different, however, and we show that $\brackett{C}$ overestimates the concentration in the tails. We illustrate these results using numerical simulations of the alternating sine flow. 

The differences in tail behaviour take a crucial importance for reacting scalars. When a logistic term is added to \eqref{eq:C_advdiff}, leading to an FKPP model, the scalar dynamics is characterised by reaction fronts whose speed can be determined from the large-deviation rate function characterising the dispersion of the non-reacting scalar \cite{Gartner1979,Freidlin1985}. We demonstrate the value of this observation in the context of flows that are random in time by predicting the location of a chemical front propagating in the alternating sine flow and checking the prediction against a  numerical simulation. 

The paper is organised as follows. Section \ref{sec:AltSineFlow} motivates the problem considered by introducing the random alternating sine flow and presenting numerical results which illustrate its action on a passive scalar. Section \ref{sec:Diffusiveregime} establishes the equivalence of $C$ and $\brackett{C}$ in the diffusive approximation that applies to the core of the scalar concentration when the spatial average of ${\bm u}$ vanishes. We check this prediction in the case of the alternating sine flow for which the effective diffusivity can be computed analytically by exploiting its ensemble-average interpretation. We consider the large-deviation aspects in section \ref{sec:LD_Regime}. We show there how the single-realisation rate function can be obtained by computing the largest Lyapunov exponent of a family of partial differential equations with coefficients which are random-in-time and periodic in space. We carry out this computation for the alternating sine flow and show that the single-realisation rate function so obtained differs from the ensemble-average one (which is given in closed form up to a Legendre transform). We highlight the importance of our findings for reacting scalars by examining front propagation in the FKPP model with the random alternating sine flow in section \ref{sec:FKPP}. We conclude in section \ref{sec:Conc}.

\section{\label{sec:AltSineFlow}A simple model of random dispersion: the alternating sine model}

To make the problem concrete, we start by introducing a simple model of random dispersion. We consider the 2D alternating sine flow with random phases popularised by \citet{Perrehumbert1994} and still widely studied (see e.g. \cite{Haynes2005,Meunier2010}, among others). This is one of the simplest random flows that efficiently stretches and folds material lines, making it uniformly mixing. This feature is crucial to enhance dispersion over that achieved by the sole action of molecular diffusion. 
The model simplifies both analytical and numerical computations by treating advection and diffusion in alternate steps in a manner akin to standard time-splitting numerical methods \cite{valocchi1992}.  

The velocity field of this  alternating sine model is spatially periodic and random in time. The scalar field obeys the evolution equation 
\begin{equation}\label{eq:SplitAlterSineModel}\!\!
\partial_{t}C=
\begin{cases}
-4a\sin\left(y+\varphi_n\right)\partial_{x}C&\text{for } t\in\left[n,n+1/4\right)\\
-4a\sin\left(x+\psi_n\right)   \partial_{y}C&\text{for } t\in\left[n+1/4,n+1/2\right)\\
2\kappa\nabla^{2}C                          &\text{for } t\in\left[n+1/2,n+1\right)
\end{cases}
\end{equation}
where $n\in\mathbb{N}$ labels the iterations,  $a$ is the maximum excursion distance in $x$ and $y$ over an iteration, $\left(\varphi_n,\psi_n\right)$ are random phases drawn independently from a uniform distribution in $\left[0,2\pi\right]$, and $\kappa$ is the molecular diffusivity. 
In \eqref{eq:SplitAlterSineModel}, time and space have been non-dimensionalised to set the time of each iteration -- in effect the correlation time of the velocity -- to 1 and the spatial period to $2\pi$.

The time-splitting enables a direct stepwise integration, turning the dynamics into that of a map. Let us introduce $C_{n}\left(\bm{x}\right)=C\left(\bm{x},t=n\right)$, with $\bm{x}=(x,y)$.
The solutions of the first two steps in \eqref{eq:SplitAlterSineModel} are obtained as  $C(x,y,t)=C_{n}(x-4a\sin(y+\varphi_n)(t-n),y)$ and $C(x,y,t)=C_{n+\frac{1}{4}}(x,y-4a\sin(x+\psi_n)(t-n-1/4))$ by integrating the advection equations. The solution of the diffusion equation in the third step is simply a convolution of $C_{n+\frac{1}{2}}(\bm{x})$ with a Gaussian diffusion kernel. This gives the 3-step map
\begin{subequations}\label{eq:stepmodel}
\begin{align} 
    C_{n+\frac{1}{4}}\left(x,y\right) &= C_{n}\left(x-a\sin\left(y+\varphi_n\right),y\right),\label{eq:AdvX}\\
    C_{n+\frac{1}{2}}\left(x,y\right) &= C_{n+\frac{1}{4}}\left(x,y-a\sin\left(x+\psi_n\right)\right),\label{eq:AdvY}\\ 
    C_{n+1}&=\mathcal{K}*C_{n+\frac{1}{2}},\label{eq:Diff}
\end{align}
\end{subequations}
where $\mathcal{K}(\bm{x})=(4\pi\kappa)^{-1}\ee^{-|\bm{x}|^2/(4\kappa)}$  and $*$ denotes the convolution product. Numerically, the iterations \eqref{eq:stepmodel} are carried out by discretising $C$ over a uniform square grid, with the  two advection steps \eqref{eq:AdvX}--\eqref{eq:AdvY} performed using a bi-linear interpolation and the convolution of the diffusion step \eqref{eq:Diff} performed in Fourier space.

\begin{figure}
    \centering
    \includegraphics[width=0.6\linewidth]{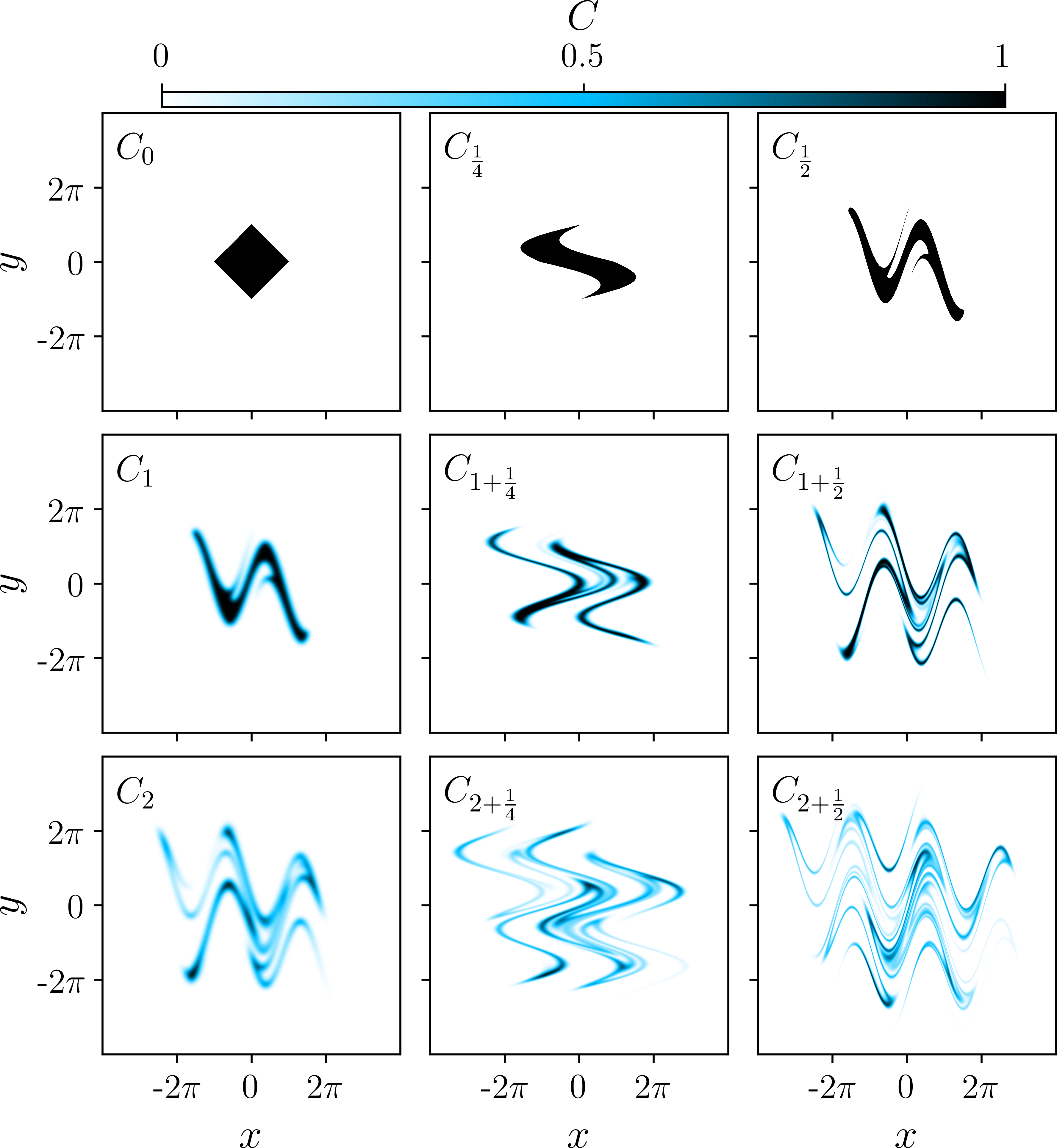}
    \caption{Evolution of an initial diamond\--shaped scalar patch (upper\--right panel) through the successive steps (\ref{eq:stepmodel}a-c) of the alternating flow with $a=\pi$ and $\kappa=10^{-2}$. The computation uses a $1024^2$ discretisation of the domain $[-4\pi,4\pi]^2$.}
    \label{fig:SineFlow}
\end{figure}

\begin{figure}
    \centering
    \includegraphics[width=0.6\linewidth]{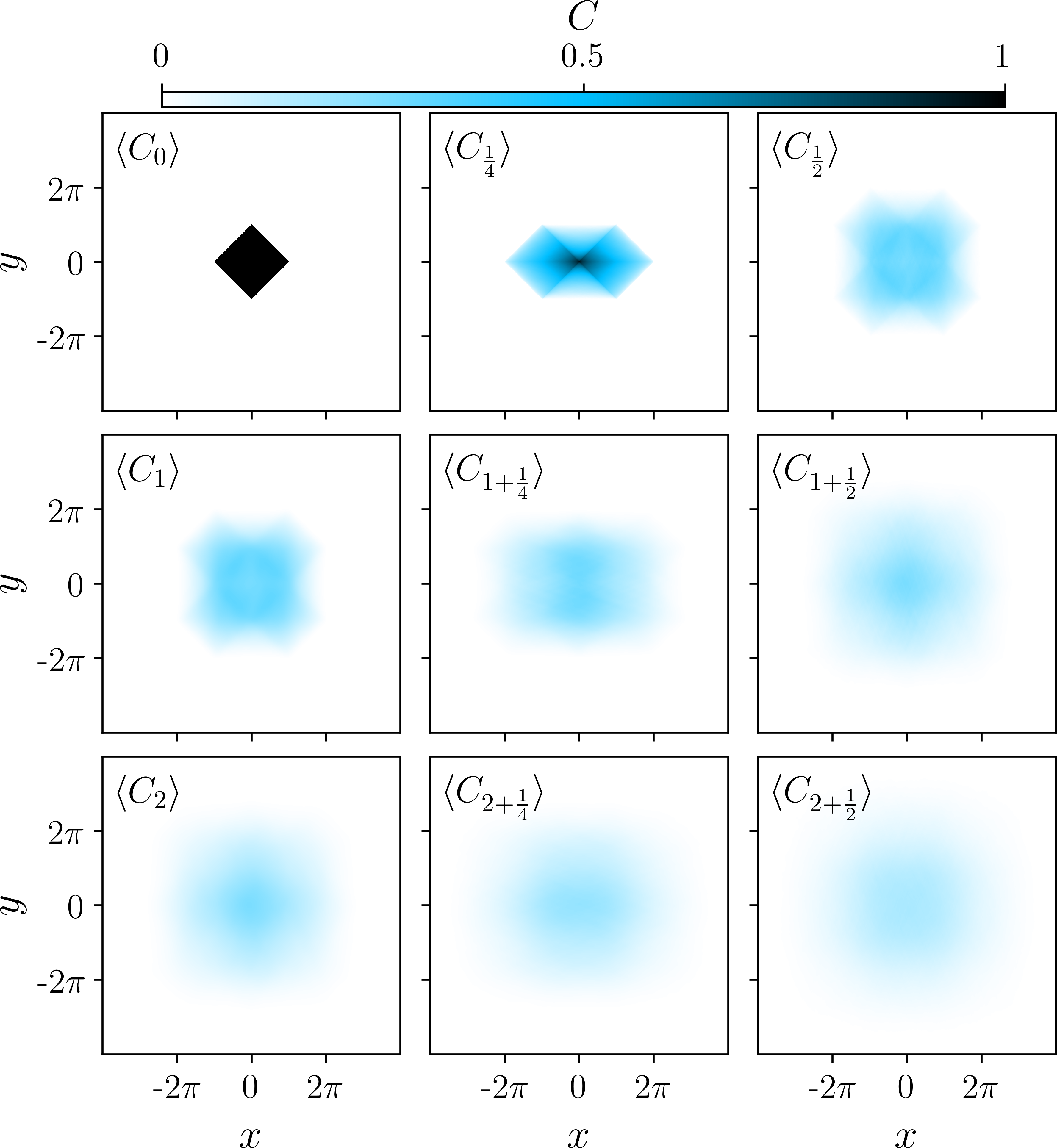}
    \caption{Same as figure \ref{fig:SineFlow} but with an average over $1000$ realisations of the flow.}
    \label{fig:Ensemble}
\end{figure}

Figure \ref{fig:SineFlow} illustrates the dynamics \eqref{eq:stepmodel} by showing the concentration field $C_n$ and the intermediate fields $C_{n+\frac{1}{4}}$ and $C_{n+\frac{1}{2}}$ for $n=0,1,2$, starting from a diamond\--shaped scalar patch. It can be contrasted with figure \ref{fig:Ensemble} which shows the same fields but averaged over $1000$ realisations of the velocity field. The single-realisations and ensemble-averaged concentration are quite different in these early iterates of the alternating sine flow, but
our focus is on the long-time (large-$n$) behaviour when the scalar has spread over many flow periods and is well represented by a coarse-grained description. In this limit,  $C_n$ and $\brackett{C_n}$ are related in a manner that we elucidate in the rest of the paper.

We emphasise that we consider  scalar dynamics in an unbounded domain. In a bounded domain, achieved e.g.\ by imposing periodic boundary conditions for $C_n$, the evolution is radically different, with a scalar relaxing exponentially fast towards a homogeneous state and adopting the form of a strange eigenmode \cite{Perrehumbert1994,Tsang2005,HaynesVanneste2005}.

\section{\label{sec:Diffusiveregime}Diffusive approximation}

In the long-time limit, when the passive scalar has sampled the velocity field well, the advection--diffusion equation \eqref{eq:C_advdiff} is often  approximated by a diffusion equation
\begin{equation}
    \partial_t C + {\bm{v}} \cdot \nabla C = \nabla \cdot \left( \mathrm{K} \cdot \nabla C \right),
\end{equation}
where ${\bm{v}}$ is a uniform effective velocity and $\mathrm{K}$ a uniform effective-diffusivity tensor \citep{majda1999,PavliotisStuart2007}. Correspondingly, the concentration field takes a Gaussian form. This is a manifestation of the central limit theorem which applies at distances that are $O(\sqrt{t})$ away from the centre of mass. The diffusive approximation implies the asymptotic linear growth with time of the mean and covariance of the position of particles obeying \eqref{eq:SDE}. The effective velocity and diffusivity
\begin{subequations}
    \begin{align}
     \bm{v}&= \lim_{t\to\infty}\frac{\mathbb{E}\left[\bm{X}\left(t\right)\right]}{t}, \\ 
        \mathrm{K}&=\lim_{t\to\infty}\frac{\mathbb{E}\left[\bm{X}\left(t\right)\otimes\bm{X}\left(t\right)\right]-\mathbb{E}\left[\bm{X}\left(t\right)\right]\otimes\mathbb{E}\left[\bm{X}\left(t\right)\right]}{2t}, \label{eq:diffK}
    \end{align}
\end{subequations}
where $\otimes$ denotes the tensor product,
then characterise the dispersion for a specific realisation of the velocity field. We expect $\bm{v}$ and $\mathrm{K}$ to be identical for almost all realisations of the velocity field: this is the property of self-averaging which arises provided that the ensemble of velocity field is well sampled in the long-time limit. When if holds, it is clear that $\bm{v} = \brackett{\bm{v}}$ and $\mathrm{K} = \brackett{\mathrm{K}}$.

The property of self-averaging has been explored, mostly for media that are random in space; though delicate to establish rigorously, it appears to be generic in the absence of anomalous diffusion \cite[e.g.][]{Bouchaud1990}. In section \ref{sec:LD_Regime} we give an argument for flows that are random in time which indicates that self-averaging holds generically not only for $\bm{v}$ and $\mathrm{K}$ but more generally for all the cumulants of $C$ divided by $t$. Numerical results demonstrate this to apply to the alternating sine model.

The ensemble averaged concentration $\bracket{C}$ is also approximately Gaussian, with corresponding effective velocity and diffusivity defined as
\begin{subequations}
    \begin{align}
     \overline{\bm{v}}&\equiv\lim_{t\to\infty}\frac{\brackett{\mathbb{E}\left[\bm{X}\left(t\right)\right]}}{t}, \\ 
        \overline{\mathrm{K}}&\equiv\lim_{t\to\infty}\frac{\brackett{\mathbb{E}\left[\bm{X}\left(t\right)\otimes\bm{X}\left(t\right)\right]}-\brackett{\mathbb{E}\left[\bm{X}\left(t\right)\right]}\otimes\brackett{\mathbb{E}\left[\bm{X}\left(t\right)\right]}}{2t}.  \label{eq:diffKbar}
    \end{align}
\end{subequations}
With the self averaging property, we readily obtain that $\bm{v}=\overline{\bm{v}}$ and 
\begin{equation}
    {\rm K}=\overline{\rm K}+\lim_{t\to\infty}\frac{\brackett{\mathbb{E}\left[\bm{X}\left(t\right)\right]}\otimes\brackett{\mathbb{E}\left[\bm{X}\left(t\right)\right]}-\brackett{\mathbb{E}\left[\bm{X}\left(t\right)\right]\otimes\mathbb{E}\left[\bm{X}\left(t\right)\right]}}{2t}.
    \label{eq:aa}
\end{equation}
 It is not obvious whether $\mathrm{K}=\overline{\mathrm{K}}$ or not. The difference depends on the size of the fluctuations in the position of the centre of mass $\mathbb{E}\left[\bm{X}\left(t\right)\right]$ between realisations of the flow (see \cite{LeDoussal1989,Bouchaud1990} for a discussion in the context of spatially random media]).
 In the next subsections, we establish  that the equality holds for spatially periodic flows  provided that the spatial average of $\bm{u}$ vanishes, and we verify this conclusion for the alternating sine model. This equivalence between single-realisation and ensemble-averaged diagnostics of dispersion does not generalise to high-order cumulants as we discuss in section \ref{sec:LD_Regime}.

\subsection{\label{ssec:DiffEq} Equivalence for spatially periodic random flows with vanishing mean}

We restrict our attention to  flows $\bm{u}$ that are random in time, with stationary statistics, zero mean, and finite correlation time, and $2\pi$-periodic in space with vanishing spatial average. We show that flows in this class satisfy $\mathrm{K}=\overline{\rm K}$, hence that 
the single-realisation and ensemble-averaged scalar fields $C$ and $\bracket{C}$ are equivalent in the diffusive approximation. 
We start by averaging (\ref{eq:SDE}) with respect to both the Brownian motion and the velocity realisation. This gives $\brackett{\mathbb{E}\left[\bm{\bm{X}\left(t\right)}\right]}=\bm{0}$, since $\brackett{\bm{u}(\cdot,t)}=0$, hence
\begin{equation} 
\bm{v}=\overline{\bm{v}}=\bm{0}
\end{equation}
and 
\begin{equation}
    \mathrm{K}=\overline{\rm K}-\lim_{t\to\infty}\frac{\brackett{\mathbb{E}\left[\bm{X}\left(t\right)\right]\otimes\mathbb{E}\left[\bm{X}\left(t\right)\right]}}{2t}
\end{equation}
using \eqref{eq:aa}.
We now show that $\left\|\brackett{\mathbb{E}\left[\bm{X}\left(t\right)\right]\otimes\mathbb{E}\left[\bm{X}\left(t\right)\right]}\right\|=o\left(t\right)$.

In order to estimate $\mathbb{E}\left[\bm{X}\left(t\right)\right]$, we introduce the expected displacement at time $t$ of a particle initially located at $\bm{x}$,
\begin{equation}\label{eq:DisplExp}
    \bchi\left(\bm{x},t\right)=\mathbb{E}\left[\bm{X}\left(t\right)\left|\bm{X}\left(0\right)=\bm{x} \right.\right]-\bm{x},
\end{equation}
such that $\mathbb{E}\left[\bm{X}\left(t\right)\right]=\bchi\left(0,t\right)$. The expectation in \eqref{eq:DisplExp} satisfies the backward Kolmogorov equation \citep[e.g.][Ch.\ 8]{okse98}
\begin{equation}
 \partial_{t}(\bchi+\bm{x})-\left(\bm{u}\cdot\nabla\right)(\bchi+\bm{x})=\kappa\Delta(\bchi+\bm{x}),
\end{equation}
hence $\bchi$ satisfies the corresponding forced equation
\begin{equation}\label{eq:BckwKolmog}
    \partial_{t}\bchi-\left(\bm{u}\cdot\nabla\right)\bchi=\kappa\Delta\bchi+\bm{u},
\end{equation} 
with the initial condition $\bchi\left(\bm{x},0\right)=\bm{0}$. 
Since $\bm{u}$ and the initial condition $\bchi\left(\bm{x},0\right)$ are both spatially periodic, the solution $\bchi$ of Eq.\ (\ref{eq:BckwKolmog}) is spatially periodic. The cell-integrated variance
\begin{equation}\label{eq:defq}
    \sigma\left(t\right)=\bracket{\mathcal{A}\left[ |\bchi|^{2}\left(\bm{x},t\right)\right]},
\end{equation}
where $\mathcal{A}$ denotes spatial averaging over a periodic cell, is found from \eqref{eq:BckwKolmog} to satisfy
\begin{equation} \label{eq:sigma}
    \frac{\mathrm{d}\sigma}{\mathrm{d}t}=-2\kappa\bracket{\mathcal{A}\left[\|\|\nabla\bchi\|\|^2\right]}+2\bracket{\mathcal{A}\left[\bm{u}\cdot\bchi\right]}
\end{equation}
using integration by parts and the incompressibility condition $\nabla \cdot \bm{u}=0$ and where $\|\|.\|\|$ denotes the Frobenius norm.
We bound the right-hand side of \eqref{eq:sigma} in terms of $\sigma$. Averaging Eq.\ (\ref{eq:BckwKolmog}) and using that $\mathcal{A}\left[\bm{u}\right]=0$ gives $\mathcal{A}\left[\bchi\right]=0$. This enables us to use the Poincaré inequality for the term $\mathcal{A}\left[\|\|\nabla\bchi\|\|^2\right]$ and the Cauchy--Schwarz inequality for the term $\mathcal{A}\left[\bm{u}\cdot\bchi\right]$. This yields the differential inequality
\begin{equation}\label{eq:ineqq}
    \frac{\mathrm{d}\sigma}{\mathrm{d}t}\leq-2\kappa \sigma+2U\sqrt{\sigma},
\end{equation}
where the domain is taken as $[0,2\pi]^d$ so that the Poincaré constant is $1$ and $U=\brackett{\mathcal{A}[|{\bm u}|^2]}^{1/2}$   is the root-mean-square velocity.
Integrating in time, we obtain
\begin{equation}\label{eq:ineqFinal}
    \sigma\left(t\right)\leq\frac{U^{2}}{\kappa^{2}}\left(1-\ee^{-\kappa t}\right)^2\leq\frac{U^2}{\kappa^2}.
\end{equation}
The Laplacian operator in Eq.\ (\ref{eq:BckwKolmog}) ensures that the solution $\bchi$ is smooth, so the boundedness of $\sigma$ guaranteed by \eqref{eq:ineqFinal} implies that $\bracket{|\bchi|^{2}\left(0,t\right)}$ is also bounded. 
Finally, observing that $\left\|\brackett{\mathbb{E}\left[\bm{X}\left(t\right)\right]\otimes\mathbb{E}\left[\bm{X}\left(t\right)\right]}\right\|=O\left(\bracket{|\bchi|^{2}\left(0,t\right)} \right)=O(1)=o\left(t\right)$ we conclude that $\mathrm{K}=\overline{\mathrm{K}}$.

We note that the upper bound $U^2/\kappa^2$ found in \eqref{eq:ineqFinal} is far from optimal. We also note that the argument above relies crucially on the assumption $\mathcal{A}\left[\bm{ u}\right]={\bf 0}$. Without this premise, it is possible that $\mathrm{K} \neq  \overline{\mathrm{K}}$. As an example, consider the case of a spatially uniform 2D flow with constant magnitude $U$ whose direction changes randomly every unit time. Such a flow has a non\--zero spatial average but its ensemble average vanishes. Computing the two diffusivity tensors is straightforward and gives  $\mathrm{K}=\kappa\mathbb{I}_d$ and $\overline{\mathrm{K}}=\left(\kappa+U/4\right)\mathbb{I}_d$, where $\mathbb{I}_{d}$ is the $d \times d$ identity matrix. 

\subsection{\label{ssec:diffregCbraket} Application to the alternating sine model}

In this subsection, we compare the diffusive approximation of $C_n$ and $\bracket{C_n}$ for the  alternating sine model introduced in section \ref{sec:AltSineFlow}. We introduce the finite\--time velocities
\begin{equation}\label{eq:FT_drift}
    \bm{v}_n\equiv\frac{\mathbb{E}\left[\bm{X}_n\right]}{n}\quad\text{,}\quad\overline{\bm{v}}_n\equiv\frac{\brackett{\mathbb{E}\left[\bm{X}_n\right]}}{n},
\end{equation}
and the finite\--time diffusivity tensors
\begin{subequations}\label{eq:FT_Diff}
    \begin{align}
        \mathrm{K}_n\equiv&\frac{\mathbb{E}\left[\bm{X}_n\otimes\bm{X}_n\right]-\mathbb{E}\left[\bm{X}_n\right]\otimes\mathbb{E}\left[\bm{X}_n\right]}{2n}\label{eq:FT_Diff-1},\\
        \overline{\mathrm{K}}_n\equiv&\frac{\brackett{\mathbb{E}\left[\bm{X}_n\otimes\bm{X}_n\right]}-\brackett{\mathbb{E}\left[\bm{X}_n\right]}\otimes\brackett{\mathbb{E}\left[\bm{X}_n\right]}}{2n},
    \end{align}
\end{subequations}
which converge to $\bm{v}$, $\overline{\bm{v}}$, $\mathrm{K}$ and $\overline{\rm K}$ in the large\--$n$ limit.

The discrete-time equivalent of \eqref{eq:SDE} associated with the map \eqref{eq:stepmodel} is the  random walk for the position $\bm{X}_{n}=\left(X_{n},Y_{n}\right)$ of a single scalar particle,
\begin{subequations}\label{eq:randomwalk}
\begin{align} 
    \bm{X}_{n+\frac{1}{4}}& = \bm{X}_{n}+ a\sin\left(Y_{n}+\varphi_n\right)\bm{e}_{x},\label{eq:RW_AdvX}\\
    \bm{X}_{n+\frac{1}{2}}&= \bm{X}_{n+\frac{1}{4}}+a\sin(X_{n+\frac{1}{4}}+\psi_n)\bm{e}_{y},\label{eq:RW_AdvY}\\ 
     \bm{X}_{n+1}&=\bm{X}_{n+\frac{1}{2}}+\sqrt{2\kappa} \bm{W}_{n},\label{eq:RW_Diff}
\end{align}
\end{subequations}
where the $\bm{W}_{n}$ are independent Gaussian variables with zero mean and unit variance. Starting from $\bm{X}_0=\bm{0}$, the scalar field $C_n$ satisfying  \eqref{eq:stepmodel}, once suitably normalised, is the probability density function 
\begin{equation}\label{eq:XtoC}
    C_n\left(\bm{x}\right)=\mathbb{E}\left[\delta\left(\bm{X}_{n}-\bm{x}\right)\right],
\end{equation}
of the particle position, with $\mathbb{E}$ denoting the expectation with respect to the Gaussian variables $\bm{W}_{n}$ associated to the molecular diffusion. Taking ${\bm X}_0={\bm 0}$ corresponds to a Dirac initial condition  $C_{0}\left(\bm{x}\right)=\delta\left(\bm{x}\right)$, which we employ in what follows.

For the alternating sine flow, each realisation is defined by a sequence of phases $(\varphi_n,\psi_n)$, so the ensemble average is 
\begin{equation}\label{eq:EnsAverage}
    \bracket{C_{n}}=\frac{1}{\left(2\pi\right)^{2n}}\underset{\left(\boldsymbol{\varphi},\boldsymbol{\psi}\right)\in\left[0,2\pi\right]^{n}}{\idotsint}\mathrm{d}\boldsymbol{\varphi}\mathrm{d}\boldsymbol{\psi}\,C_{n},
\end{equation}
where $\boldsymbol{\varphi}=\left(\varphi_{0},\dots,\varphi_{n-1}\right)$ and $\boldsymbol{\psi}=\left(\psi_{0},\dots,\psi_{n-1}\right)$.
We note that
\begin{equation}\label{eq:AdvX_av}
\left\langle C_{n}\right.^{^{\!\!\!\!\!\!(1)}}\left.\left(x,y\right)\right\rangle = \bracket{C_{n}\left(x-a\sin\varphi_n,y\right)},
\end{equation}
which is readily established by the substitution $\phi_n \mapsto \phi_n - y$ in the integral \eqref{eq:EnsAverage} defining the ensemble average. Similarly, 
\begin{equation}\label{eq:AdvY_av}
\left\langle C_{n}\right.^{^{\!\!\!\!\!\!(2)}}\left.\left(x,y\right)\right\rangle = \left\langle C_{n}\right.^{^{\!\!\!\!\!\!(1)}}\!\!\!\left.\left(x,y-a\sin\psi_n\right)\right\rangle.
\end{equation}
As a result, $\bracket{C_{n}}$ can be obtained as the ensemble-averaged probability density function
\begin{equation}\label{eq:SimpelRel_C_Bracket}
    \bracket{C_n\left(\bm{x}\right)}=\bracket{\mathbb{E}\left[\delta\left(\tilde{\bm{X}}_n-\bm{x}\right)\right]}
\end{equation}
associated with the modified random walk
\begin{subequations}\label{eq:randomwalkbar}
\begin{align} 
    \tilde{\bm{X}}_{n+\frac{1}{4}}& = \tilde{\bm{X}}_{n}+ a\sin \varphi_n \bm{e}_{x},\label{eq:RWB_AdvX}\\
    \tilde{\bm{X}}_{n+\frac{1}{2}}&= \tilde{\bm{X}}_{n+\frac{1}{4}}+a\sin \psi_n \bm{e}_{y},\label{eq:RWB_AdvY}\\ 
     \tilde{\bm{X}}_{n+1}&=\tilde{\bm{X}}_{n+\frac{1}{2}}+\sqrt{2\kappa} \bm{W}_{n}. \label{eq:RWB_Diff}
\end{align}
\end{subequations}
This has the advantage over \eqref{eq:randomwalk} that the position after $n$ steps can be written explicitly as
\begin{equation}\label{eq:effectiveDyn}
    \tilde{\bm{X}}_{n}= a \sum_{k=0}^{n-1}\left(\sin\varphi_k\bm{e}_{x}+\sin\psi_k\bm{e}_{y}\right)+\sqrt{2\kappa n}\bm{W},
\end{equation}
where $\bm{W}$ is a single Gaussian variable with zero mean and unit variance.

\begin{figure}
    \centering
    \includegraphics[width=0.5\linewidth]{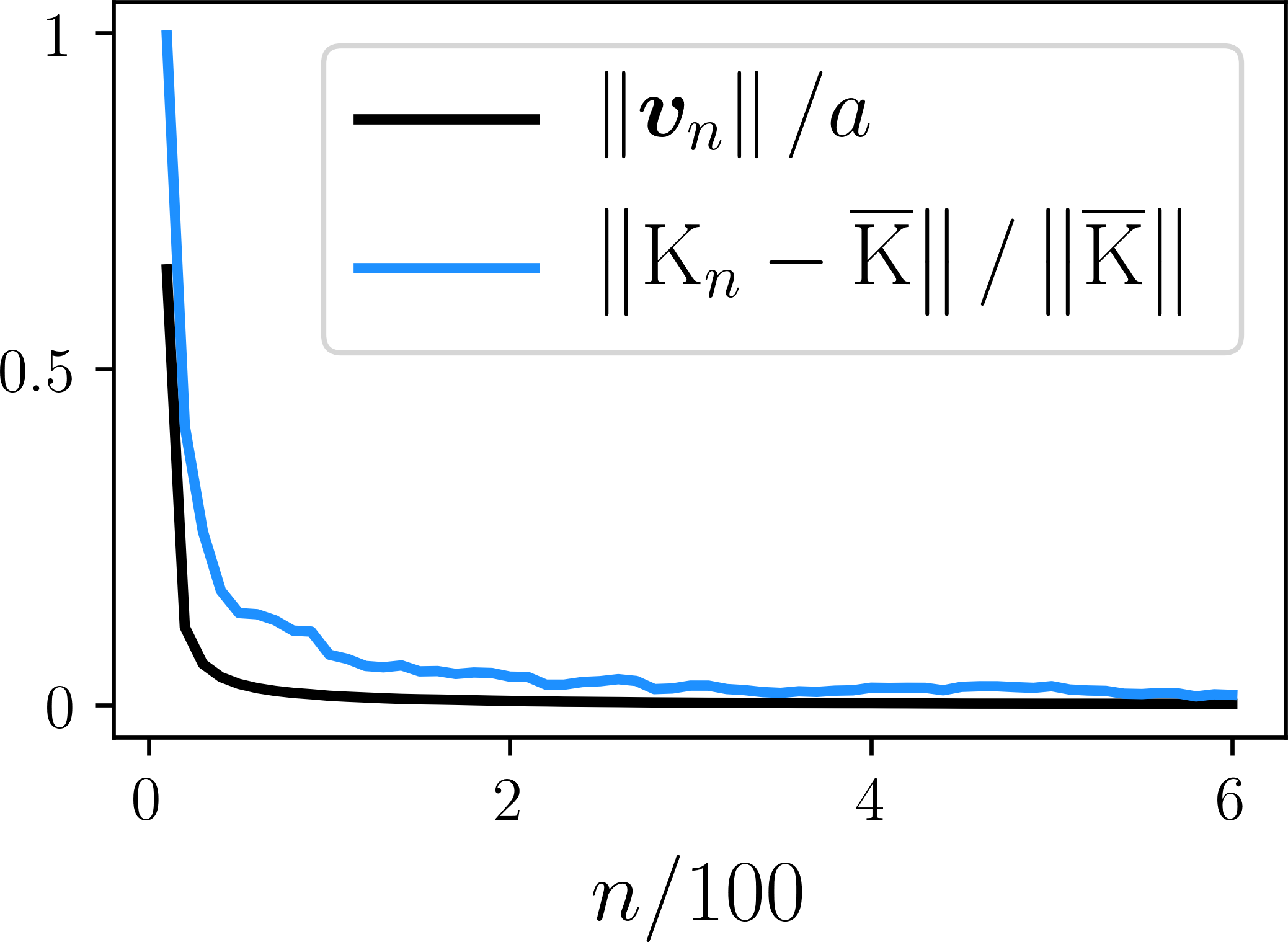}
    \caption{Magnitude of the effective velocity ${\bm v}_n$ and of the (relative) difference between the effective diffusivity ${\rm K}_n$ and the ensemble-average  $\overline{\rm K} = \left(\kappa+a^2/4\right)\mathbb{I}_2$ as functions of $n$ for the alternating sine flow for $a=\pi$ and $\kappa=10^{-2}$. 
    Here, $\bm{v}_n$ and ${\rm K}_n$ are defined in Eqs.\ (\ref{eq:FT_drift}) and (\ref{eq:FT_Diff-1})  and  estimated by numerical sampling of the random walk (\ref{eq:randomwalk}) using $10^4$ realisations of the Brownian motion.}
    \label{fig:DiffEq}
\end{figure}

Replacing ${\bm X}_n$ by $\tilde{\bm X}_n$ leaves the ensemble-averaged statistics unchanged, as Eq.\ \eqref{eq:SimpelRel_C_Bracket} shows, while making their computation straightforward. We can in particular compute $\overline{\bm v}_n$ and $\overline{\rm K}_n$ by substituting $\tilde{\bm{X}}_{n}$ for ${\bm X}_n$ in \eqref{eq:FT_drift} and \eqref{eq:FT_Diff} to obtain
\begin{equation}\label{eq:cbarkbar}
    \overline{\bm{v}}_n= \overline{\bm{v}}=\bm{0}\quad\text{and}\quad
    \overline{\mathrm{K}}_{\,n}=\overline{\mathrm{K}}=\left(\kappa+\frac{a^{2}}{4}\right)\mathbb{I}_{2}.
\end{equation}
An argument similar to that in section  \ref{ssec:DiffEq} but adapted the discrete-time setup can be used to show that ${\bm v} = \overline{\bm v}$ and $\mathrm{K}=\overline{\mathrm{K}}$ so that $C_n$ and $\bracket{C_n}$ shares the same diffusive approximation. We do not reproduce this argument here. Instead, we confirm its conclusion numerically by sampling the random walk \eqref{eq:randomwalk} numerically, estimating 
$\bm{v}_n$ and $\mathrm{K}_n$, and verifying their convergence towards $\overline{\bm v}$ and $\overline{\mathrm{K}}$ in \eqref{eq:cbarkbar} as shown in Figure \ref{fig:DiffEq}. This illustrates the  equivalence between single-realisation and ensemble-averaged predictions in the diffusive approximation when only the first two cumulants matter. We next investigate the behaviour of higher-order cumulants by considering the large-deviation approximation of the concentration.

%%%%%%%%%%%%%%%%%%%%%%%%%%%%%%%%%%
%%%%%%%%%%%%%%%%%%%%%%%%%%%%%%%%%%
\section{\label{sec:LD_Regime}  Large\--deviation approximation}

\subsection{\label{ssec:Formulation_LDReg} General formulation}

The effective velocity ${\bm v}$ and effective diffusivity $\mathrm{K}$ characterise only the core of the scalar distribution, occupying a $O(\sqrt{t})$ range around the centre of mass. At larger distances, the concentration field is generally not Gaussian; instead, it takes the large-deviation form
\begin{equation} \label{eq:largeDeviation}
    C(\bm{x},t) \asymp \ee^{-t g(\bm{x}/t)},
\end{equation}
valid for $|\bm{x}-\bm{v} t| = O(t)$. Here $g$ is the rate function (or Cramér function) and the symbol $\asymp$ denotes the asymptotic equivalence of the logarithms of the expressions on both sides (this ignores a constant multiple of $t^{-d/2}$ which can be determined by mass conservation). The approximation \eqref{eq:largeDeviation} holds for almost all realisations of the flow ${\bm u}$ for a broad class of flows.
The Gaussian approximation is recovered from \eqref{eq:largeDeviation} by Taylor expanding $g$ around its minimum, $\bm{\xi}_*$ say, to find
\begin{equation}
    \bm{v} = \bm{\xi}_* \quad \textrm{and} \quad \mathrm{K} = \tfrac{1}{2} \left(\nabla \nabla g(\bm{\xi_*})\right)^{-1}. 
\end{equation}

Eq.\ \eqref{eq:largeDeviation} goes beyond this and captures all the cumulants of the scalar function. Indeed, it can be shown that the scaled  (single-realisation) cumulant generating function, 
\begin{equation} \label{eq:f}
f(\bm{q}) \equiv \lim_{t\to\infty} t^{-1}  \log  \mathbb{E} \ee^{\bm{q} \cdot \bm{X}(t)} 
\end{equation}
is the Legendre transform of the rate function \cite{HaynesVanneste2014}: 
\begin{equation}
    f(\bm{q}) = \inf_{\bm{\xi}} \left( \bm{q} \cdot \bm{\xi} - g(\bm{\xi})) \right). 
\end{equation}
The Legendre duality implies that $g$ is the Legendre transform of $f$. This provides a means to evaluate $g$ since $f$ can be computed as the Lyapunov exponent of a random-in-time partial differential equation. To see this, we note that the function
\begin{equation} \label{eq:wDef}
    w(\bm{x},\bm{q},t) \equiv \mathbb{E} \left[ \ee^{\bm{q} \cdot \bm{X}(t)} | \bm{X}(0)=\bm{x} \right],
\end{equation}
satisfies the backward Kolmogorov equation \citep[e.g.][Ch.\ 8]{okse98}
\begin{equation} \label{eq:backKolmogorov}
    \partial_t w =  \bm{u} \cdot \nabla w + \kappa \nabla^2 w.
\end{equation}
In view of the initial condition $w(\bm{x},\bm{q},0)=\exp(\bm{q} \cdot \bm{x})$ and the spatial periodicity of ${\bm u}$, the solution of \eqref{eq:backKolmogorov} can be sought in form 
\begin{equation} \label{eq:w}
w(\bm{x},\bm{q},t) = \exp\left(\bm{q} \cdot \bm{x} \right) \phi(\bm{x},\bm{q},t),
\end{equation} 
where $\phi(\bm{x},\bm{q},t)$ has the same spatial periodicity as $\bm{u}$. Introducing this into Eq.\ \eqref{eq:backKolmogorov}, $\phi$ is found to satisfy
\begin{equation} \label{eq:phi}
    \partial_t \phi = \kappa\nabla^2 \phi + \left(\bm{u} + 2 \kappa \bm{q} \right) \cdot \nabla \phi + \left(\bm{u} \cdot \bm{q} + \kappa |\bm{q}|^2 \right) \phi.
\end{equation}
In the limit $t \to \infty$, $\phi$ grows exponentially with $t$ at a rate -- the Lyapunov exponent of \eqref{eq:phi} --  that can be identified from Eqs.\ \eqref{eq:f}, \eqref{eq:wDef} and \eqref{eq:w} as $f(\bm{q})$. From the theory of Lyapunov exponents in finite \cite{osel68} and infinite \cite{ruel82} dimensions, $f(\bm{q})$ can be expected to be independent of the realisation of the flow under some assumptions of ergodicity.
Thus we can obtain the rate function $g$ governing the large-deviation statistics of the scalar concentration in (almost all) single realisations of the velocity field by solving \eqref{eq:phi} for $\phi$ with periodic boundary conditions and an arbitrary initial condition for a range of values of $\bm{q}$, identifying $f(\bm{q})$ as the growth rate of $\phi$, i.e., as
\begin{equation}
f(\bm{q}) = \lim_{t \to \infty} t^{-1} \log \| \phi \| 
\end{equation}
for some norm $\| \cdot \|$, then carrying out a numerical Legendre transform to deduce $g$. We note that the independence of $f(\bm{q})$ on flow realisation implies that the scaled cumulant generating function and hence all the cumulants of $C$ themselves, when divided by $t$, are independent of the flow realisation; in other words, that they are self-averaging.

The large-deviation characterisation of the concentration $C(\bm{x},t)$ in single realisations of the velocity has a counterpart for the ensemble-averaged concentration $\brackett{C(\bm{x},t)}$. Specifically, there is an ensemble-average large-deviation rate function $\overline{g}$ such that
\begin{equation}
\brackett{C(\bm{x},t)} \asymp \ee^{-t \overline{g}(\bm{x}/t)}.
\end{equation}
Its Legendre dual is the ensemble-average cumulant generating function 
\begin{equation} \label{eq:fff}
\overline{f}(\bm{q}) \equiv \lim_{t \to \infty} t^{-1} \log \brackett{\mathbb{E}\, \ee^{\bm{q} \cdot \bm{X}(t)}}.
\end{equation} 
Note that its single-realisation counterpart can be rewritten as
\begin{equation} \label{eq:overlinefff}
   f(\bm{q}) = \lim_{t\to\infty} t^{-1} \brackett{ \log  \mathbb{E} \, \ee^{\bm{q} \cdot \bm{X}(t)}},  
\end{equation}
on ensemble averaging Eq.\ \eqref{eq:f} since \eqref{eq:f} applies to almost all realisations of ${\bm u}$. Thus the difference between the two generating functions stems from the lack of commutation between ensemble average and logarithm. 
Note that the definitions \eqref{eq:fff}--\eqref{eq:overlinefff}, the concavity of the log and Jensen's inequality imply that $f \le \overline{f}$, which guarantees that 
\begin{equation} \label{eq:inequality}
g \ge \overline{g}
\end{equation}
using the order-reversing property of the Legendre transform. Thus, the ensemble-averaged scalar field always overestimates the tail  concentration in single realisations of the flow at large time.

A focus of this paper is the relationship between  $f$ and $\overline{f}$ and, correspondingly, between $g$ and $\overline{g}$ 
The equalities $\bm{v} = \overline{\bm{v}}$ and $\mathrm{K}=\overline{\mathrm{K}}$ established in the previous sections imply that Taylor expansions at $\bm{\xi}=\bm{\xi}_*$ of $g$ and $\bar g$ are identical up to and including quadratic terms. The same is true for $f$ and $\overline{f}$ by virtue of the Legendre duality with $g$ and $\overline{g}$. The equalities do not however hold for higher order terms: in general, $g \not= \overline{g}$ and $f \not= \overline{f}$. Thus the single-realisation and ensemble-averaged concentration fields, which are identical in the diffusive approximation, differ, leading to differences in the cumulants of order greater than 2. Physically, this means that the tails of the concentration in single realisations,  unlike the core, cannot be predicted on the basis of ensemble-averaged statistics. We next demonstrate this explicitly for the  alternating sine flow.

\subsection{\label{ssec:SASmodel_LD} Application to the alternating sine model}

For the  alternating sine flow, we use the discrete version of the large-deviation rate functions $g$ and $\overline{g}$, such that
\begin{equation}\label{eq:discrete_LD}
    C_n \asymp \ee^{-n g({\bm x}/n)} \quad \textrm{and} \quad \brackett{C_n} \asymp \ee^{-n \overline{g}({\bm x}/n)},
\end{equation}
and 
\begin{equation}
     {f}\left(\bm{q}\right) = \lim_{n\to\infty}\frac{1}{n} \brackett{\log\mathbb{E} \ee^{\bm{q}\cdot\bm{X}_n}},
     \quad \textrm{and} \quad 
        \overline{f}\left(\bm{q}\right) = \lim_{n\to\infty}\frac{1}{n}\log\brackett{\mathbb{E} \ee^{\bm{q}\cdot\bm{X}_n} }\label{eq:fCndef-1}.
\end{equation}
The ensemble-averaged $\overline{f}$ can be computed explicitly, replacing the positions $\bm{X}_n$ by   $\tilde{\bm{X}}_n$ in Eq.\ \eqref{eq:effectiveDyn}, exploiting their statistical equivalence \eqref{eq:SimpelRel_C_Bracket}.  This gives
\begin{subequations}
\begin{align}\label{eq:fbar}
    \overline{f}({\bm q})&= \frac{1}{n} \log  \left(\sum_{k=0}^{n-1} \bracket{\ee^{a(q_x \sin \varphi_k + q_y \sin \psi_k)}} + \mathbb{E} \ee^{\sqrt{2\kappa n}{\bm q} \cdot {\bm W}} \right)  \\
    &=\log {I}_{0}\left(aq_x\right) +\log {I}_0\left(aq_y\right)+\kappa |\bm{q}|^{2},
\end{align}
\end{subequations}
where ${\bm q}=(q_x,q_y)$, on using the integral definition of the modified  Bessel function of the first kind $I_{0}$ \cite[Eq.~10.32.1]{NIST:DLMF}. Note that a Taylor expansion using that $I_0(a q_x) \sim 1 - a^2 q_x^2/4 $ as $q_x \to 0$ and similarly for $I_0(a q_y)$ recovers the diffusive approximation  \eqref{eq:cbarkbar}.

Obtaining the single-realisation $f$ is not as straightforward: this requires estimating numerically the Lyapunov exponent of the discrete-in-time equivalent of \eqref{eq:phi}, namely
\begin{equation}\label{eq:Discrete_Eigenval}\!\!
\partial_{t}\phi=
\begin{cases}
4a\sin\left(y+\varphi_n\right)(q_x+\partial_{x})\phi&\text{for } t\in\left[n,n+1/4\right)\\
4a\sin\left(x+\psi_n\right)   (q_y+\partial_{y})\phi&\text{for } t\in\left[n+1/4,n+1/2\right)\\
2\kappa(\nabla^2+2\bm{q}\cdot\nabla+|\bm{q}|^2)\phi                          &\text{for } t\in\left[n+1/2,n+1\right)
\end{cases}.
\end{equation}
This can be integrated in time to obtain the 3-step discrete map
\begin{subequations} \label{eq:eigenvalsteps}
\begin{align} 
    \phi_{n+\frac{1}{4}}(x,y)&=\phi_n(x+a\sin(y+\varphi_n),y)\ee^{q_xa\sin(y+\varphi_n)} \label{eq:eigenvalstep1}\\
    \phi_{n+\frac{1}{2}}(x,y)&=\phi_{n+\frac{1}{4}}(x,y+a\sin(x+\psi_n))\ee^{q_ya\sin(x+\psi_n)} \label{eq:eigenvalstep2} \\
    \phi_{n+1}&=\widetilde{\mathcal{K}}*\phi_{n+\frac{1}{2}} \label{eq:eigenvalstep3}
\end{align}
\end{subequations}
where $\phi_n(x,y)=\phi(x,y,n)$, $\widetilde{\mathcal{K}}(\bm{x})=(4\pi\kappa)^{-1}\ee^{\bm{q}\cdot\bm{x}-|\bm{x}|^2/(4\kappa)}$  and $*$ denotes the convolution product.

For a given $\bm{q}$, we estimate $f(\bm{q})$ by iterating \eqref{eq:eigenvalsteps}  numerically  for $n$ up to $10^5$, starting from $\phi_0(x,y)=1$. The function $\phi_n(x,y)$ is discretised on a $128^2$ uniform spatial grid. The first two steps, \eqref{eq:eigenvalstep1}--\eqref{eq:eigenvalstep2}, are performed in a semi-Lagrangian fashion, as  a circular permutation of the spatial grid indices; the third step, \eqref{eq:eigenvalstep3}, is performed spectrally, in Fourier space. The growth rate $f(\bm{q})$ is approximated as $n^{-1}\log\|\phi_n\|$ for $n$ large enough that the approximation has converged. The convergence is rather slow, with $n=10^5$ iterations required here. We carry out this numerical estimation of $f(\bm{q})$ for $100^2$ values of $(q_x,q_y)$ linearly spaced in $[0,1]^2$, using symmetries to obtain $f({\bm q})$ in $[-1,1]^2$.  

\begin{figure}
    \centering
    \includegraphics[width=0.85\linewidth]{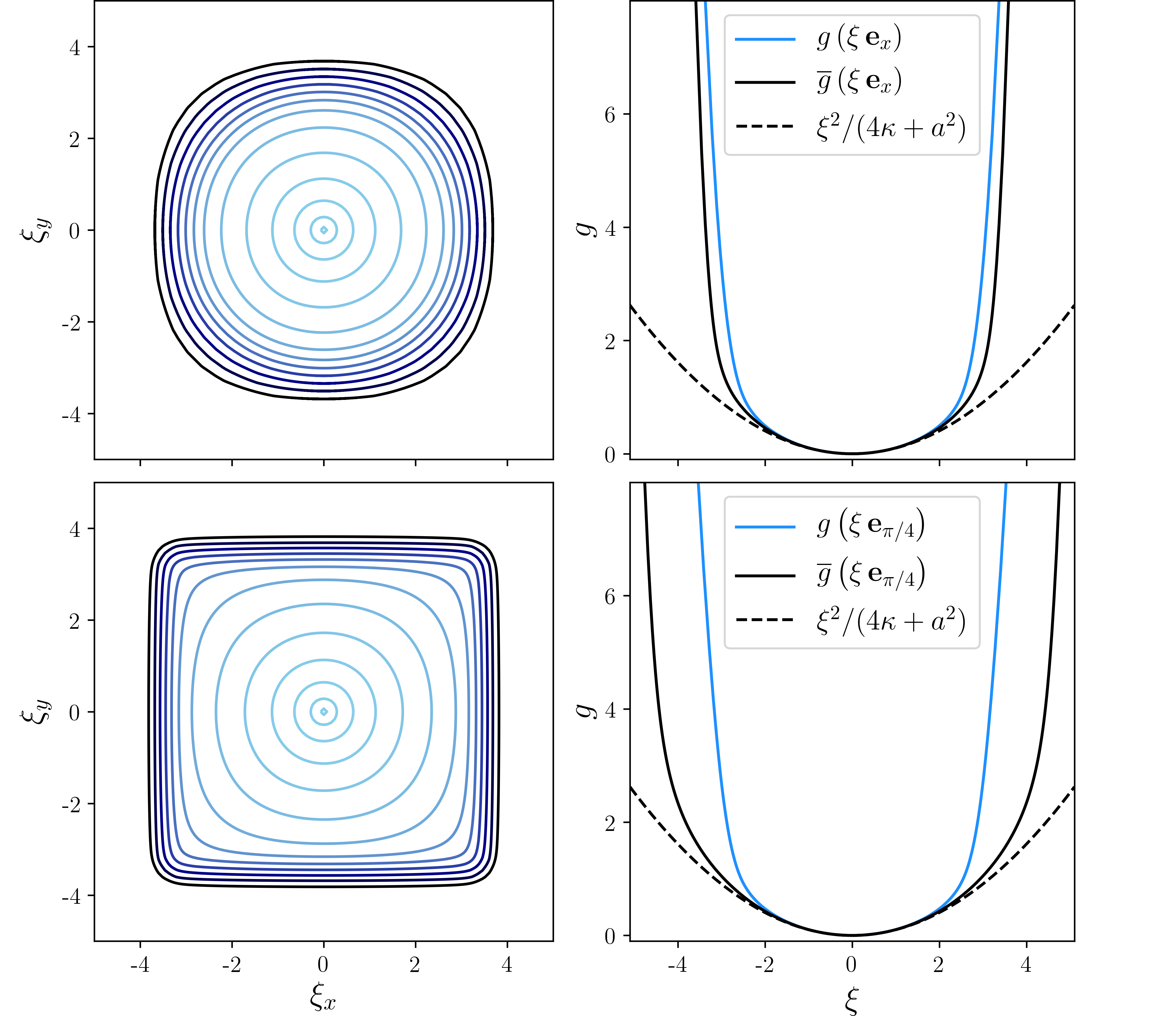}
    \caption{Comparison of the single-realisation and ensemble-average large\--deviation rate functions $g$ and $\overline{g}$ for the alternating sine model with $a=\pi$ and $\kappa=10^{-2}$. Left panels: contours of $g$ (top) and $\overline{g}$ (bottom) (using the same contour levels for both panels). Right panels:  $g$ and $\overline{g}$ along the $x$-axis (top) and along a diagonal axis (bottom). The diffusive approximation is also shown (dashed).}
    \label{fig:Xi_g}
\end{figure}

We show the results for the rate functions $g$ and $\overline{g}$, obtained from $f$ and $\overline{f}$ by numerical Legendre transform, in Fig.\  \ref{fig:Xi_g}. The two functions are markedly different away from the centre ${\bm \xi} = {\bm \xi}_* = {\bm 0}$ where the diffusion approximation applies and around which they are approximately equal. The function $g$ grows substantially faster than $\overline{g}$ (cf.\ Eq.\ \eqref{eq:inequality}). Consequently, the ensemble-averaged $\brackett{C_n}$ considerably overestimates the tail concentration in single realisations of the flow. The departure from axis-symmetry, present for both $g$ and $\overline{g}$, is the result of the flow geometry: fast dispersion stems from particles experiencing maximum advection at each step, which can lead to the maximum displacements $(\pm a,\pm a)$ at each time step (ignoring diffusion). This indicates that for sufficiently large ${\bm \xi}$ (and sufficiently small $\kappa$), the contours of $g$ and $\overline{g}$ are approximately square as would be the case for advective displacements randomly chosen from $(\pm a,\pm a)$ at each iteration \cite[cf.][]{TzellaVanneste2015a}. This effect is much more marked for the ensemble-averaged concentration than for  single realisations. 

The broad conclusion is that the ensemble-averaged concentration is a poor predictor of the behaviour of dispersing passive scalars when it comes to their low-concentration tails. While the tail behaviour might seem of marginal interest, it plays a crucial role in the presence of chemical reactions, in particular by determining the dynamics of some reaction fronts. In this instance, the differences between single-realisation and ensemble-averaged dynamics can be dramatic as we discuss next.

\section{\label{sec:FKPP}Application to reactive front propagation}

\subsection{FKPP model}

We now consider the dynamics of a reacting scalar governed by the classical FKPP model of Fisher \cite{Fisher1937} and Kolmogorov \emph{et al.} \cite{Kolmogorov1937} which, in the presence of advection, reads
\begin{equation}\label{eq:FKPP}
    \partial_t C +\bm{u}\cdot\nabla C= \kappa \nabla^2 C + \gamma C(1-C),
\end{equation}
where $\gamma>0$ is the reaction rate and the concentration $C$ has been normalised so that its saturation value be $1$. According to this model, in an infinite domain, an initially localised patch of scalar spreads behind a reaction front. 

For ${\bm u}=0$ and for $t \gg 1$, the front is circular in two dimensions (spherical in three) with a radius growing at the constant speed $s = 2 \sqrt{\kappa \gamma}$. For a spatially periodic flow, the front can be approximated by a self-similar convex curve (or surface) defined by 
\begin{equation}\label{eq:front_eq}
    g\left(\frac{\bm{x}}{t}\right)=\gamma,
\end{equation}
where $g$ is the large-deviation rate function introduced in \eqref{eq:largeDeviation} for the passive scalar problem \eqref{eq:C_advdiff} (i.e.\ $\gamma=0$ in \eqref{eq:FKPP}) \citep{Gartner1979,Freidlin1985,Xin2000b,Tzella2014,TzellaVanneste2015b}. The remarkable connection between the large-deviation tail of a passive scalar  and the FKPP reaction front  embodied by \eqref{eq:front_eq} can be understood heuristically as follows. The front's propagation is controlled by its leading edge, where $C \ll 1$ so that \eqref{eq:FKPP} can be linearised about $C=0$; the solution is then that of the passive-scalar problem \eqref{eq:C_advdiff}, approximately $\exp\left(-t g({\bm x}/t)\right)$, multiplied by the factor $\exp(\gamma t)$ associated with the (linearised) reaction, hence 
\begin{equation}
    C({\bm x},t) \asymp \ee^{t\left(\gamma - g({\bm x}/t) \right)}.
\end{equation}
Eq.\ \eqref{eq:front_eq} follows by observing that the front marks the transition between exponential growth and exponential decay. 

For the random-in-time flows considered in this paper, $g$ is clearly the single-realisation rate function, so the shape of the front can be determined from \eqref{eq:front_eq} using that $g$ is the Legendre transform of the Lyapunov $f$ exponent of the random partial differential equation \eqref{eq:phi}. The ensemble-averaged $\overline{g}$, while easier to evaluate, is only relevant to the front problem in the limit of small reaction rate $\gamma \ll 1$, when the diffusive approximation applies, $g \approx \overline{g}$, and the front is the ellipse (or ellipsoid)
\begin{equation}
    ({\bm x} - {\bm v} t)^{\mathrm{T}} {\rm K}^{-1}  ({\bm x} - {\bm v} t) = 4 \gamma t^2
\end{equation}
as follows from Taylor expanding \eqref{eq:front_eq}.

\subsection{Reacting alternating sine model} \label{sec:reactingsine}

We illustrate how \eqref{eq:front_eq} predicts the shape of a reacting front by considering a reacting version of the alternating sine model \eqref{eq:stepmodel}. This adds the reaction step
\begin{equation} \label{eq:reactionstep}
    C_{n+1}=C_{n+\frac{3}{4}}/(C_{n+\frac{3}{4}}+(1-C_{n+\frac{3}{4}})\ee^{-\gamma}),
\end{equation}
where $C_{n+\frac{3}{4}}=\mathcal{K}*C_{n+\frac{1}{2}}$, to the advective and diffusive steps in \eqref{eq:stepmodel}, corresponding to the time-$1$ map associated with the FKPP reaction $\partial_t
C = \gamma C(1-C)$. The front location is determined by the discrete-time equivalent of \eqref{eq:front_eq}, that is,
\begin{equation}\label{eq:SAS_front}
    g\left(\frac{\bm{x}}{n}\right)=\gamma.
\end{equation}
Note that this equation makes it possible to interpret the contours of $g$ in Fig.\ \ref{fig:Xi_g} as giving the shape of fronts for different values of $\gamma$. 

\begin{figure}
    \centering
    \includegraphics[width=\linewidth]{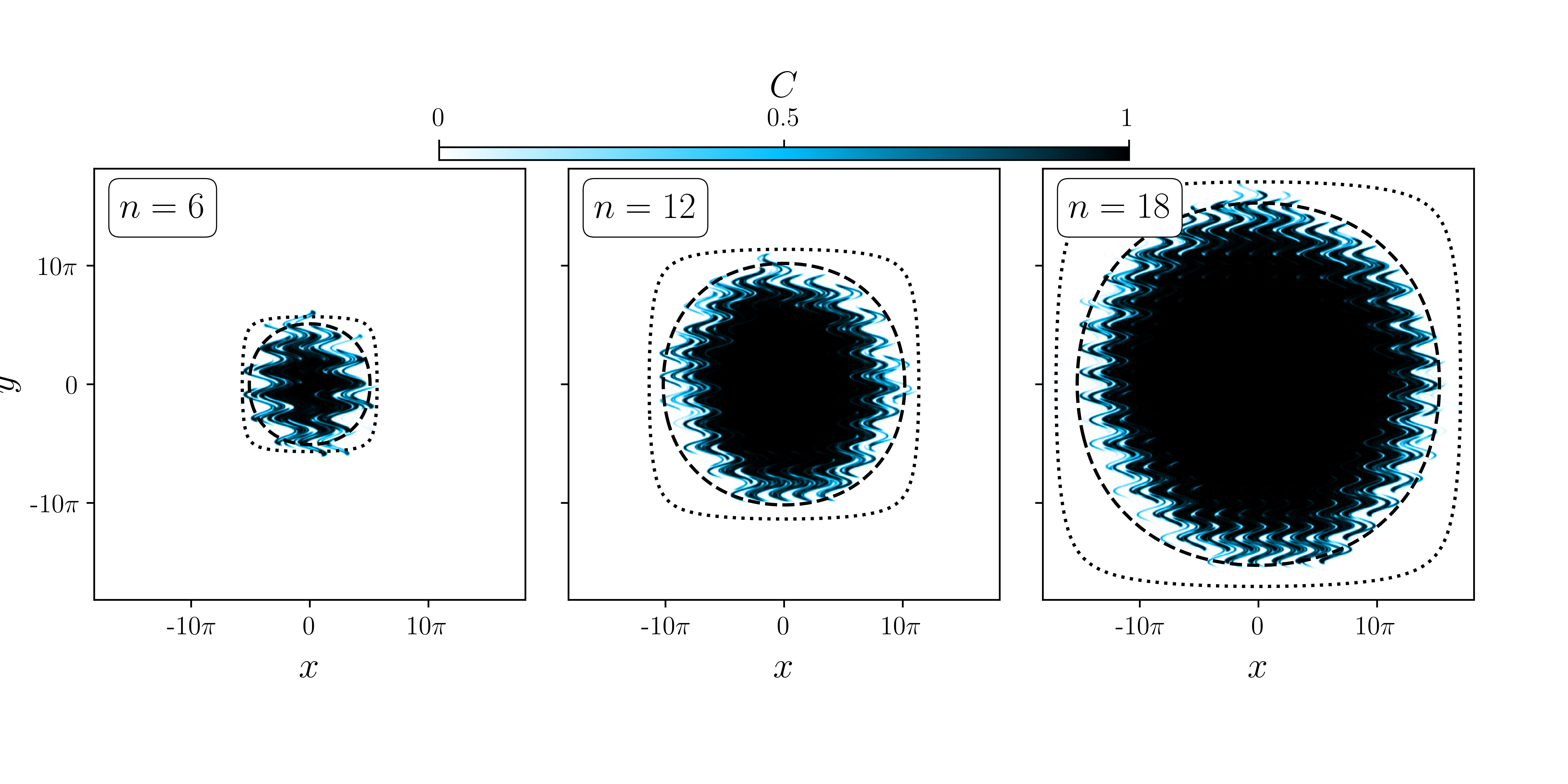}
    \caption{Evolution of the concentration for the reacting alternating sine model in \S\ref{sec:reactingsine} with $a=\pi$, $\kappa=0.1$ and $\gamma=2$. The prediction of the front location according to \eqref{eq:SAS_front}, with the single-realisation $g$ estimated as in \S\ref{ssec:SASmodel_LD} is shown as the solid blue line; its counterpart using the ensemble averaged $\overline{g}$ is shown as the dashed black line. The computation uses a $4096^2$ discretisation of a $[-64\pi,64\pi]^2$ domain.}
    \label{fig:FKPP}
\end{figure}

Figure \ref{fig:FKPP} shows the concentration field obtained by direct numerical integration of the reacting alternating sine model starting from a localised patch with $a=\pi$, $\kappa=0.1$ and $\gamma = 2$. The result is compared with the prediction \eqref{eq:SAS_front} for the front location. The prediction which substitutes the ensemble averaged $\overline{g}$ (computed in subsection \ref{ssec:SASmodel_LD}) is also shown. The figure confirms the validity of   \eqref{eq:SAS_front} and illustrates the inadequacy of the prediction based on $\overline{g}$.  

\section{\label{sec:Conc}Summary and conclusion}
This paper investigates the large-time statistics of passive scalars dispersing under the combined action of molecular diffusion and advection by single realisations of flows taken to be spatially periodic and random-in-time. The focus is on the instantaneous release of a localised patch of scalar in an unbounded domain. After some time, the scalar patch extends over many flow periods and has sampled flow correlation time such that it is well represented using a coarse-grained description, and the  velocity $\bm{v}$ and diffusivity $\mathrm{K}$ are identical in almost all flow realisations (that is, self-averaging applies). This happens for times large compared to both the correlation time of the flow, and the typical time for dispersion across a periodic cell which depends on both molecular diffusivity value and  flow strength.

The single-realisation statistics are compared to their ensemble-averaged counterparts which are more easily accessible to analysis and prominently studied in the literature. Under the assumption of a vanishing spatial average of the flow, we show that the Gaussian cores of the ensemble-average and single-realisation scalar distributions are identical. 
We introduce a large-deviation description which captures the tails of the concentration, hence higher-order cumulants,  and show that these are in general overestimated by the ensemble-average prediction. In the single-realisation problem, we demonstrate that the large-deviation rate function can be obtained by computing the largest Lyapunov exponent for the spatially periodic solution of a family of partial differential equations with coefficients that are random-in-time. We illustrate these results 
by computing the single-realisation rate function for the widely-used alternating sine flow and comparing it with its ensemble-average counterpart which is obtained analytically.

The tail behaviour of the concentration of a dispersing passive scalar has a zeroth-order manifestation in the presence of a chemical reaction of FKPP type. This is because the large-time dynamics of the reaction front that is then formed is directly related to the large-deviation rate function characterising the dispersion. We consider this problem which further highlights the importance of the distinction between ensemble-average and single-realisation predictions for the scalar-concentration tails of the scalar distributions. For the random alternating sine flow, we successfully predict the location of the reaction front using the single-realisation rate function computed for the non-reacting scalar. The corresponding prediction based on the ensemble-average rate function fails. 

In conclusion, we emphasise the existence of a hierarchy of behaviours for scalars dispersing in random flows: at the most basic level, the self-averaging of the cumulants scaled by $t^{-1}$ ($\bm{v}$, $\mathrm{K}$ and higher order) can be expected to hold for typical flows. Next, the equivalence between single-realisation and ensemble-average statistics in a diffusive approximation (that is, the equalities $\bm{v}=\overline{\bm{v}}$ and $\mathrm{K}=\overline{\mathrm{K}}$) is more delicate but applies to periodic flows with vanishing spatial average. We expect it to also hold for spatially random flows with finite correlation length and vanishing spatial average. Finally, the analogous equivalence for higher-order cumulants, which characterise the large-deviation approximation, does not hold, even for such a well-behaved model as the alternating sine flow. We also note that the ensemble average overestimates the value of the scaled cumulants of order 3 and higher; in physical terms, it overestimates the concentration in the tails of the scalar distribution.

We finally observe that the distinction between single-realisation and ensemble-average predictions of scalar concentrations is an issue of broad interest, arising whenever stochastic models are used to represent complex velocity fields. While ensemble-averaged statistics are usually more amenable to analysis, single-realisation statistics are often more meaningful for practical applications such as risk assessment. In view of this broad interest, it would be desirable to extend the analysis of the present paper to other classes of flows, and in particular to flows that are random in space \cite{isishenkoreview}. We also note that the problem posed here in unbounded domains has a counterpart in bounded domains, in which case the focus will be on the differences between the rate of decay of the scalar concentration in single realisation and in ensemble averages \cite{HaynesVanneste2014}. 

\begin{acknowledgments}
This research is funded by grant EP/S023291/1 of the UK Engineering and Physical Sciences Research Council. We thank Alexandra Tzella and the two anonymous referees for useful comments.
\end{acknowledgments}

\bibliography{SineFlows}

\end{document}